\documentclass[a4paper,prd,11pt,noshowkeys,notitlepage,nofootinbib]{revtex4}
\pdfoutput=1  
\usepackage[latin1]{inputenc}
\usepackage{ae,aecompl}
\usepackage{amsmath,amssymb,mathrsfs}
\usepackage{graphicx}
\usepackage{dsfont}
\usepackage{slashed}
\usepackage{units}

\usepackage[usenames,dvipsnames]{xcolor} 
\usepackage{hyperref}
\hypersetup{
    colorlinks=true,       
    linkcolor=purple,          
    citecolor=ProcessBlue,        
    filecolor=magenta,      
    urlcolor=YellowOrange           
}

\providecommand{\cM}{\mathscr{M}}
\providecommand{\la}[1]{\lambda_{#1}}

\providecommand{\dst}{\displaystyle}

\providecommand{\btheta}{\bar{\theta}}
\providecommand{\Vckm}{V_{\rm CKM}}
\providecommand{\cp}{\mathsf{CP}}
\providecommand{\hx}{\hat{x}}

\providecommand{\msbar}{$\overline{\mathrm{MS}}$}
\providecommand{\lag}{\mathscr{L}}
\providecommand{\cD}{\mathcal{D}}
\providecommand{\eps}{\epsilon}

\makeatletter
\newcommand*\bigcdot{\mathpalette\bigcdot@{.5}}
\newcommand*\bigcdot@[2]{\mathbin{\vcenter{\hbox{\scalebox{#2}{$\m@th#1\bullet$}}}}}
\makeatother

\providecommand{\eq}[1]{\begin{equation} #1 \end{equation}}
\providecommand{\eqali}[1]{\begin{equation}\begin{aligned} #1
    \end{aligned}\end{equation}}
\providecommand{\subeqali}[2][]{\begin{subequations}\label{#1}\begin{align}
#2    \end{align}\end{subequations}}    
\providecommand{\ZZ}{\mathbb{Z}}
\DeclareMathOperator{\re}{\mathrm{Re}} 
\DeclareMathOperator{\im}{\mathrm{Im}} 
\providecommand{\mtrx}[1]{\begin{pmatrix} #1 \end{pmatrix}}
\providecommand{\mss}[1]{\mbox{\scriptsize $#1$}}
\providecommand{\tp}{{\mss{\mathsf{T}}}}
\providecommand{\ums}[2][1]{\ml{\tfrac{#1}{#2}}} 
\providecommand{\ml}[1]{\mbox{\large $#1$}}
\providecommand{\aver}[1]{\langle #1 \rangle}
\DeclareMathOperator{\diag}{\mathrm{diag}} 
\usepackage{bbm}
\providecommand{\id}{{\mathbbm{1}}} 
\DeclareMathOperator{\Tr}{\mathrm{Tr}} 
\providecommand{\xlink}[1]
  {\href{http://arxiv.org/abs/#1}{arXiv:#1}}

\begin{document}
\title{
Solving the strong CP problem with non-conventional CP
}
\author{A.~L.~Cherchiglia}
\email{adriano.cherchiglia@ufabc.edu.br}
\author{C.~C.~Nishi}
\email{celso.nishi@ufabc.edu.br}
\affiliation{Universidade Federal do ABC - UFABC, Santo Andr\'{e}, SP, Brasil}

\begin{abstract}
A very simple model is presented where all CP violation in Nature is spontaneous in origin.
The CKM phase is generated unsuppressed and the strong CP problem is solved with only moderately small couplings between the SM and the CP violation sector or mediator sector because corrections to $\btheta$ arise only at two loops.
The latter feature follows from an underlying unconventional CP symmetry of order 4 imposed in the sectors beyond the SM composed of only two vector-like quarks of charge $-1/3$ and one complex scalar singlet.
No additional symmetry is necessary to implement the Nelson-Barr mechanism.
\end{abstract}
\maketitle
\section{Introduction}
\label{sec:intro}

The fact that Nature distinguishes left from right and particles from antiparticles at energies so far explored was established long ago and represents a cornerstone of our ability to infer the most basic properties of the fundamental interactions.
In its most subtle form this symmetry violation, called CP violation, has been experimentally confirmed only through the presence of one irremovable phase in the mixing matrix among quarks interacting with the $W$ bosons in the weak interactions, a manifestation known as the CKM mechanism.
In principle, the nontrivial vacuum structure of QCD would allow CP violation to appear through the nonzero value of the so-called $\btheta$ parameter, with possible contamination from phases in the weak sector.
Why this parameter is experimentally constrained to be so small is known as the strong CP problem
(see \cite{Kim:2008hd} for a review).

The explanation for the strong CP problem usually invokes three possibilities: (i) 
the promotion of $\btheta$ to a dynamical field -- the axion -- which couples to the QCD gluon potential and then is dynamically driven to zero in the potential minimum\,\cite{PQ};
(ii) a massless up quark which allows $\btheta$ to be rotated away by the global axial symmetry (strongly disfavoured by lattice calculations\,\cite{lattice});
(iii) CP (or P) is indeed conserved at the fundamental level and its violation manifests itself only through spontaneous breaking at lower energies making $\btheta$ calculable and to arise only at loop level, potentially justifying its tiny value\,\cite{spontaneous,nelson,barr}.
Within approach (iii), the Nelson-Barr mechanism is one of the simplest ways to guarantee $\btheta=0$ at tree-level from explicit CP conservation\,\cite{nelson,barr}.
For other approaches consisting on P conservation, see Ref.\,\cite{P-vio}.
Other proposals based on CP conservation can be seen in Ref.\,\cite{scpv:others,cp-texture}.

In the simplest implementation of the Nelson-Barr idea, Bento, Branco and Parada (BBP)\,\cite{BBP} enlarged the SM with only one complex singlet scalar and one vector-like quark of charge $-1/3$.
The former is responsible for the spontaneous breaking of CP while the latter mediates this breaking to the SM.
They have found that a correction to $\btheta$ was generated at one-loop as
\eq{
\delta\btheta\sim \frac{f^2\lambda_{\phi S}}{16\pi^2}\,,
}
where $f$ and $\lambda_{\phi S}$ quantify the two portals to the SM: the former 
is the Yukawa coupling between the heavy quarks transmitting the CP violation from the scalar sector to the SM and $\lambda_{\phi S}$ is the Higgs portal coupling(s) to the CP violating scalar.
So, sufficiently suppressed $\btheta$ required very suppressed portal couplings.
See Ref.\,\cite{dine} for more discussions on the naturality of this and similar schemes.

Here we will improve on this simple model by assuming the presence of a new order 4 CP symmetry, dubbed CP4\,\cite{cp4:inert}, acting on the new scalar and the heavy quark sector which now requires two vector-like quarks.
This symmetry further protects $\btheta=0$ which is now corrected only at two loops as
\eq{
\delta\btheta\sim \frac{f^4\lambda_{\phi S}}{(16\pi^2)^2}\,.
}
Therefore, only moderately small portal couplings are necessary to obey the current bound: $\btheta\lesssim 10^{-10}$\,\cite{EDMreview,nEDM:exp}.

The study and application of CP4 symmetry is being actively pursued in the literature after the original model was proposed in the context of a 3HDM with irremovable complex parameters in the Higgs potential without explicit CP violation\,\cite{cp4:inert}.
After that, it was extended to the quark sector in Ref.\,\cite{cp4:quarks}, where irremovable phases appear in the Yukawa sector as well, and to the neutrino sector in Ref.\,\cite{cp4:nu}.
More recently, an algorithm to detect such a symmetry in the 3HDM was devised\,\cite{cp4:inv}, its relation to mass degeneracy was also studied\,\cite{cp4:haber} and the interplay between annihilation and conversion to the DM abundance was investigated\,\cite{cp4:DM}.

We organize the rest of this paper as follows: in Sec.\,\ref{sec:cp4} we review some aspects of CP4 which will be needed to construct the model.
The model is presented in Sec.\,\ref{sec:model}.
The vanishing of $\btheta$ at one-loop is shown in Sec.\,\ref{sec:loop} together with the estimate at two-loops.
The conclusions are then shown in Sec.\,\ref{sec:conclu}.

\section{Review of CP4}
\label{sec:cp4}

Here we review the action of CP4. We begin by analyzing the action on two complex scalars which can be readily adapted to two chiral fermions.
Our model, developed in the next section, will make use of two such a pair of chiral fermions together with a single complex scalar which, as reviewed below, is equivalent to a pair of real scalars transforming faithfully under CP4.

\subsection{Scalars}

The basic structure of the order-4 CP transformation, known as CP4\,\cite{cp4:inert,cp4:quarks}, can be defined by two complex scalar fields $\varphi_1,\varphi_2$ transforming as
\eq{
\label{CP4}
\mathsf{CP4}:~~
\varphi_1(x)\to \varphi_2^*(\hat{x})\,,\quad
\varphi_2(x)\to -\varphi_1^*(\hat{x})\,,
}
where $\hat{x}$ denotes spatial inversion of $x$.
The minus sign in the second relation is crucial because its square leads to a $\ZZ_2$ symmetry:
\eq{
\label{cp4^2}
(\mathsf{CP4})^2:~~
\varphi_1\to -\varphi_1\,,\quad
\varphi_2\to -\varphi_2\,;
}
hence order 4 CP symmetry\,\cite{cp4:inert}.

Taking combinations of these fields, we can recover more familiar transformation properties. For example, we can construct CP4-odd combinations,
\eq{
\varphi_1\varphi_2^*\,,~~|\varphi_1|^2-|\varphi_2|^2\,,
}
or usual CP transformation, 
\eq{
\label{varphi1.varphi2}
\varphi_1\varphi_2\to -(\varphi_1\varphi_2)^*\,,
}
with an additional sign that can be eliminated by rephasing $\varphi_i$.

It is well known that multiplication by $i$ on a complex field $S$ can be represented by the real matrix $\eps=i\sigma_2$ in the real basis $(\re S,\im S)$.
The eigenvalues in complex space does not change, i.e., $\pm i$ for $S,S^*$.
In the same way, in the basis $(\re\varphi_1,\re\varphi_2,\im\varphi_1,\im\varphi_2)$, the transformation in \eqref{CP4} can be represented by
\eq{
\label{CP4:re.im}
\mathsf{CP4}:~~
\mtrx{-\eps & 0 \cr 0& \eps}\,.
}
This too has eigevalues $\pm i$, each one with multiplicity two. The combinations corresponding to eigenvalue $+i$ are
\eq{
\label{CP4:i}
\mathsf{CP4}\sim i:~~
\re(\varphi_1)-i\re(\varphi_2)\,,~~
\im(\varphi_1)+i\im(\varphi_2)\,.
}
Their complex conjugate have eigenvalue $-i$.

The previous ``diagonalization'' of CP4, however, is only possible for fields that do not carry other quantum numbers as separation into real and imaginary parts is necessary.
If $\varphi_1\sim\varphi_2$ carry the same $U(1)$ charge $q$, CP4 flips this charge as usual CP.
On the other hand, if $\varphi_1\sim q$ but $\varphi_2\sim -q$, then CP4 commutes with this  $U(1)$.
In fact, the entire $SU(2)$ group acting on $(\varphi_1,\varphi_2)$ commutes with CP4 as $\eps U^*=U\eps$ for any $U\in SU(2)$.

If $\varphi_1,\varphi_2$ carry no charge, CP4 do not mix the fields in \eqref{CP4:i} and only one of them can be chosen as a minimal component, i.e., a complex scalar $S$ transforming as
\eq{
\label{cp4:S}
\cp4:\quad S(x)\to -iS(\hat{x})\,,
}
hence acting as a $\ZZ_4$ transformation on field space.
In terms of its real components $S=S_1+iS_2$, they transform as
\eq{
\cp4:\quad
S_1(x)\to S_2(\hat{x})\,,\quad
S_2(x)\to -S_1(\hat{x})\,,
}
i.e., the real version of \eqref{CP4}.

As the combination \eqref{varphi1.varphi2} transforms as a complex field by usual CP transformation, we can also see that CP4 can be represented \emph{faithfully} by \eqref{CP4} but it can also be represented unfaithfully by the usual CP transformation which has order two for scalars and for this reason is sometimes denoted as CP2 to distinguish it from CP4\,\cite{cp4:quarks}.
Therefore, when we say that a model is CP4 symmetric, it means that at least one set of fields transforms faithfully as \eqref{CP4} or \eqref{cp4:S} (or equivalently for fermions) while others might transform by usual CP2 transformation, which includes the usual CP even or CP odd behaviors for singlet scalars.
These possibilities are akin to the possibility of representing $\ZZ_4$ by fields that have charges $\pm i$ (faithful) or charges $\pm 1$ (unfaithful). The latter are the nontrivial and trivial representations of the $\ZZ_2$ subgroup of $\ZZ_4$.
We disregard the possibility of CP symmetries of order higher than four\,\cite{higher.CPk}.

It is also useful to track the action of $(\cp4)^2$ in \eqref{cp4^2} which generates a $\ZZ_2$ symmetry. For the faithful representations of CP4, the action of this $\ZZ_2$ is also faithful but for the unfaithful representations, this $\ZZ_2$ acts trivially since $(\cp2)^2$ is equivalent to the identity transformation for scalars.

\subsection{Spinors}

We define the usual CP (CP2) transformation on chiral fermion fields $\psi=\psi_{L,R}$ as
\eq{
\label{cp2:psi}
\cp2:\quad \psi(x)\to \psi^{cp}(\hx)\,,
}
where
\eq{
\label{cp:psi}
\psi^{cp}\equiv i\beta \psi^c=-iC\psi^*\,,
}
with $\beta=\gamma^0$ and $C\equiv i\gamma^0\gamma^2$ in the Dirac or  Weyl representation. 
In contrast, we define the nonconventional action of CP, i.e., CP4, on a pair of chiral 
fields $\psi_1,\psi_2$, as
\eq{
\label{cp4:psi}
\cp4:\quad \psi_1(x)\to i\psi_2^{cp}(\hx)\,,\quad
\psi_2(x)\to -i\psi_1^{cp}(\hx)\,,
}
where $cp$ is the usual CP transformation in \eqref{cp:psi}.
We will often pack the two fields as one doublet $\psi=(\psi_1,\psi_2)^\tp$ which 
transforms as
\eq{
\label{cp4:psi}
\cp4:\quad \psi(x)\to i\eps\psi^{cp}(\hx)\,,
}
where $i\eps=-\sigma_2$ acts in this degenerate space.
See Refs.\,\cite{cp4:quarks,cp4:nu} for more discussions about the action of CP4 on fermions.

Again, the action of $(\cp4)^2$ is nonconventional as 
\eq{
\label{cp4^2:psi}
(\cp4)^2:\quad \psi(x)\to +\psi(x)\,,
}
while conventional CP transformation \eqref{cp2:psi} results in
\eq{
\label{cp2^2:psi}
(\cp2)^2:\quad \psi(x)\to -\psi(x)\,.
}

\section{The model}
\label{sec:model}

\subsection{Yukawa sector}

The Yukawa sector of the model coincides with the SM for charge $2/3$ quarks but is modified for charge $-1/3$ quarks by the addition of a pair of vector-like quarks 
$D_{1L},D_{1R},D_{2L},D_{2R}$ and one singlet complex scalar $S$.
Being hypercharged, each pair $(D_{1L},D_{2L})$ and $(D_{1R},D_{2R})$ can be considered as a doublet of CP4 and transform faithfully as \eqref{cp4:psi}. 
Then we simply denote these pairs as $D_L$ and $D_R$.
The scalar $S$ also transforms faithfully under CP4 as \eqref{cp4:S}.
The rest of the fields of the SM transforms by usual CP2, i.e., as in \eqref{cp:psi} for fermions or through complex conjugation for scalars.

Because only $D_{L,R}$ and $S$ transform unconventionally under $(\cp4)^2$, the Yukawa interactions for charge $-1/3$ quarks will be partly secluded into ordinary and heavy quarks as
\eq{
\label{yukawa}
-\lag_{Y} = \bar{q}_{iL} Y^d_{ij}\,\phi\,d_{jR}  + \mu_D\bar{D}_{aL} D_{aR}+ \bar{D}_{aL}\left(F_{aj} S+\bar{F}_{aj} S^*\right)d_{jR}+h.c.,
}
where $q_{iL}$ and $d_{jR}$ are SM quark fields, $i,j=1,2,3$, $a=1,2$, $Y^d$ and $\mu_D$ are real due to CP4 with $\mu_D>0$.%
\footnote{%
The term involving $\mu_D$ can be more generic, i.e., 
a $2\times 2$ matrix obeying $\eps\mu_D^*\eps^\dag=\mu_D$, 
but it can always be diagonalized with positive and equal values using appropriate $SU(2)$ reparametrizations acting on $D_{L,R}$.
In any case, $D_1,D_2$ are degenerate in mass due to CP4.
}
In addition, the Yukawa coupling $F$ is a $2\times 3$ complex matrix and its barred 
coupling is defined by
\eq{
\bar{F}\equiv \eps F^*\,,
}
with $\eps=i\sigma_2$ being the two-dimensional antisymmetric tensor.
Unlike the original Bento-Branco-Parada (BBP) model\,\cite{BBP}, here we do not need an additional $\ZZ_2$ since this symmetry is already generated by CP4.
We note that although the theory is CP conserving the coefficients $F,\bar{F}$ are intrinsically complex and cannot be transformed to real coefficients much like the CP4 symmetric 3HDM potential proposed in Ref.\,\cite{cp4:inert}.
For the latter, intrinsically complex Yukawa couplings are also present if some quarks also transform faithfully under CP4\,\cite[b]{cp4:quarks}.
The model discussed here is complementary to those in that the parameters with irremovable
phases appear exclusively in the Yukawa sector.
One may \textit{choose} $F$ to be real, in which case, $\bar{F}$ is also real and the theory is additionally invariant by CP2 for all fields. 
We discard this possibility because in this case the scalar potential is not capable of spontaneously breaking this CP2 and the CKM phase cannot be generated; see Secs.\,\ref{sec:potential} and \ref{sec:ckm}.

The most general reparametrization transformation that maintains the CP4 (and CP2) 
transformations invariant are $SU(2)$ transformations on $D_{aL},D_{aR}$ and $O(3)$ transformations on $d_{jR}$ and $q_{iL}$.
It is clear that these transformations will leave the Yukawa interactions \eqref{yukawa} form invariant.
Rephasing of $S$, except for multiples of $i$, are forbidden because the potential needs to be invariant; cf.\ Sec.\,\ref{sec:potential}.
Under a transformation of this type $F$ and $\bar{F}$ transform in the same way.
Hence, we can see from the singular value decomposition of $F$ that in the generic case not all complex phases of $F$ can be removed.

\subsection{Scalar Potential}
\label{sec:potential}

Apart from the SM Higgs doublet, we consider one complex scalar singlet under $SU(2)_L$, $\dst S=\frac{S_{1} + iS_{2}}{\sqrt{2}}$ transforming as \eqref{cp4:S} by CP4.
The most general potential invariant under CP4 is given by
\eq{
\label{potential}
V=V_{\phi} + V_{S} + V_{S\phi}\,,
}
where
\eq{
V_{\phi}=-\mu^2_\phi \phi^{\dagger}\phi + \ums{2}\lambda (\phi^{\dagger}\phi)^2
}
is the usual SM scalar potential. The next term is given by
\eq{
\label{V:S}
V_{S}=-\mu^2_S |S|^2 + \ums{2}\lambda_{1}|S|^4 - \ums{4}\lambda_{2} S^{4} -\ums{4}\lambda_{2}^{*} S^{*4}\,,
}
where we can absorb the phase of $\la{2}$ by rephasing $S$ so that we can choose $\la{2}>0$.
Finally, the interaction between $\phi$ and $S$ occurs only through the Higgs portal
\eq{
\label{V:S.phi}
V_{S\phi} = \la{\phi S}(\phi^{\dagger}\phi)S^{*}S\,.
}
We can see that the phase of $S$ only appears in the $\la{2}$ term and then the potential is minimized when $S$ is real and positive.\footnote{Note that due to CP4 there is a 4-fold degeneracy and $\arg(S)=n\pi/4$, $n=0,1,2,3,$ are all equivalent.}
Since the potential contains only one phase sensitive monomial, the manifest canonical CP symmetry cannot be broken spontaneously\,\cite{haber.surujon}. 
The CP4 symmetry, however, will be broken once $S$ gets a vev.
Compared to the original BBP model, this scalar sector has the same number of fields but less free parameters while our Yukawa sector in \eqref{yukawa} has more fields and more parameters.

At this point, we should remark that the symmetry structure of our model is crucially different from BBP-type models because it cannot be obtained from the imposition of usual CP and a $\ZZ_n$ symmetry.
The BBP model is based on $\ZZ_2$ and CP2 and it cannot forbid $S^2$ terms in the potential (hence $\btheta$ at one-loop) in contrast to our potential in \eqref{V:S} and \eqref{V:S.phi}.
The same potential as ours can be obtained by using $\ZZ_4$ but then either the $F$ or $\bar{F}$ term would be absent in the Yukawa Lagrangian \eqref{yukawa}.
With the additional imposition of CP2, $F$ or $\bar{F}$ would be real and no CP violation can be generated to account for the CKM phase; cf.\,Sec.\,\ref{sec:ckm}.
Since the physical predictions are different, we can see that CP4 is indeed a genuinely different CP symmetry that cannot be obtained from usual CP and an additional discrete symmetry.
Other examples can be seen in Refs.\,\cite{cp4:inert,cp4:quarks,cp4:nu,cp4:haber}.

Let us now show the scalar spectrum.
By defining the vevs
\eq{
\label{vevs}
\aver{S}=\frac{v_{S}}{\sqrt{2}}\,,\quad 
\aver{\phi^0}=\frac{v}{\sqrt{2}}\,,
}
with $v=246\,\unit{GeV}$ being the electroweak scale,
we can write the minimization equations as
\eqali{
-\mu^2_{\phi} + \ums{2}\lambda v^2 + \ums{2}\lambda_{\phi S} v_{S}^2 = 0\,,\\
-\mu^2_{S}+ \ums{2}\la{12} v_{S}^2 + \ums{2}\lambda_{\phi S}v^2 = 0\,,
}
which can be solved analytically. We have also defined $\la{12}\equiv\la{1}-\la{2}>0$.

After shifting the fields by their vevs, we can define $(S_0,S_1,S_2)=\sqrt{2}(\re\phi^0,\re S,\im S)$ where $S_0$ corresponds to the SM higgs direction. In this basis, the mass matrix is
\eq{
\left(\begin{array}{ccc}
\la{} v^2 & \la{\phi S}\,v v_{S} & 0 \\
\la{\phi S}\,v v_{S} & \la{12}v_{S}^2 & 0\\
0 & 0 & 2\lambda_{2}v_{S}^2
\end{array}\right).
}
The block diagonal structure is evident and follows from usual CP conservation of the potential, hence CP4 ensures CP2 for this simple potential, although CP4 is spontaneously broken. As we are going to see, this conservation of CP2 
will be crucial for the protection of $\btheta=0$ at one-loop.

We can see that the CP odd field $A=\sqrt{2}\im S$ has mass
\eq{
\label{m:A}
m_{A}^{2} = 2\lambda_{2}v_{S}^2\,,
}
while the CP even fields $s,h$ have masses
\eqali{
\label{m:s}
m_{s}^{2} &= \ums{2}[\la{12}v_{S}^2 + v^2 \lambda] + \sqrt{\la{\phi S}^2 v^2 v_{S}^2 + \ums{4}[\la{12}v_{S}^2 - v^2 \lambda]^2}
\cr
&\approx \la{12}v_S^2\,,\\
m_{h}^{2} &= \ums{2}[\la{12}v_{S}^2 + v^2 \lambda] - \sqrt{\la{\phi S}^2 v^2 v_{S}^2 + \ums{4}[\la{12}v_{S}^2 - v^2 \lambda]^2}
\cr
&\approx v^2\left[\la{}-\frac{\la{\phi S}^2}{\la{12}-\la{}v^2/v_S^2}\right]
\,,
}
where the approximate expressions are valid for $v_S\gg v$.
The lighter scalar $h$ corresponds to the 125\,GeV Higgs boson discovered in the LHC.

The mixing between the CP even scalars is given by
\eq{
\label{mixing}
\mtrx{S_0\cr S_1}=\mtrx{\cos\alpha & -\sin\alpha\cr \sin\alpha & \cos\alpha}\mtrx{h\cr s}\,,
}
where
\eq{
\tan2\alpha=-\frac{\la{\phi S}\,v/v_S}{\la{12}-\la{}v^2/v^2_S}\,.
}
Current LHC data constrains this angle to be small, satisfying $|s_\alpha|\lesssim 0.2$\,\cite{haber:mixing,singlet.SM}, which implies $t_{2\alpha}\lesssim 0.426$ and then
$v_S\gtrsim v/0.426\sim 600\,\unit{GeV}$ for order one quartic couplings.
In fact, moderately suppressed $F$ couplings will be needed for suppressed $\btheta$ and since the heavy quarks are constrained to be above the TeV scale we will typically need $v_S\gg\unit{TeV}$.

\subsection{Generating the CKM}
\label{sec:ckm}

Considering the vevs \eqref{vevs}, the down type quark mass matrix is 
\eq{
\label{mass.matrix:d,D}
\cM_{D}=\left(\begin{array}{cc}
\mu_{d} & 0 \\
M_{DS} & \mu_{D}
\end{array}\right),\quad (\mu_{d})_{ij} = \frac{v}{\sqrt{2}} Y^{d}_{ij}, \quad (M_{DS})_{aj} =  \frac{v_{S}}{\sqrt{2}}\left[F_{aj}+\bar{F}_{aj}\right]
\,.
}
For definiteness, we work in the basis in which the mass matrix for up type quarks is diagonal.
Note that $M_{DS}$ contains irremovable phases from $F$ and thus CP is spontaneously broken.
Nevertheless, $\cM_{D}$ obeys the Barr criteria\,\cite{barr,barr.2}: a complex CKM matrix can be generated but $\bar{\theta}=0$ at tree-level.%
\footnote{%
This structure might be spoiled by the dimension 5 operators $\bar{q}_L\phi D_RS$ and $\bar{q}_L\phi D_RS^*$ but they can be suppressed if the CP violation scale is low enough compared to the cutoff\,\cite{dine}.
}
We also note that $M_{DS}$ is always nonzero as $\bar{F}=-F$ implies $F=0$.
The same is true for any relation $\bar{F}=e^{i\alpha}F$.

Usual bidiagonalization allows us to write
\eq{
\label{MD}
U^{\dagger}_{L}\cM_{D}U_{R}=
    \left(\begin{array}{cc}
    \hat{M}_d & 0 \\
    0 & \hat{M}_D
    \end{array}\right),
}
with $\hat{M}_d=\diag(m_d,m_s,m_b)$ being the SM down quark masses while $\hat{M}_D=\diag(M_{\cD_1},M_{\cD_2})\gg \hat{M}_d$ corresponds to two heavy quarks.
The observable lefthanded matrix $U_{L}$ will have a hierarchical structure due to the hierarchy $v_S\gg v$ present in
\eq{
\label{MDMD}
\mathscr{M}_{D}\mathscr{M}_{D}^{\dagger}=
\left(\begin{array}{cc}
\mu_{d}\mu_{d}^{\dagger} & \mu_{d}M_{DS}^{\dagger} \\
M_{DS}\mu_{d}^{\dagger} & H_D
\end{array}\right)
\,,
}
where we have defined $H_D\equiv \mu_{D}^2+M_{DS}M_{DS}^{\dagger}$.
Such a hierarchical structure allows us to approximate
\eq{
\label{UL}
U_{L}\approx \left(\begin{array}{cc}
\id_3 & \theta_L \\
-\theta_L^{\dagger} & \id_2
\end{array}\right)\mtrx{V_{\rm CKM} & 0 \cr 0 & V_{D}},
}
with
\eq{
\theta_L=\mu_{d}M_{DS}^{\dagger}H_D^{-1}.\\
}
The first matrix in \eqref{UL} can perform the approximate block diagonalization that leads to the mass matrices for the SM down-type quarks and heavy quarks:
\subeqali[Md:MD]{
M_dM_d^\dag&\approx \mu_d(\id_3-M_{DS}^{\dagger}H_D^{-1}M_{DS})\mu_d^\dag\,,
\\
\label{MD}
M_DM_D^\dag&\approx H_D\,.
}
We can see that $M_dM_d^\dag$ contains complex phases unsuppressed by $v/v_S$ in the second term if $\mu_D\lesssim M_{DS}$, analogous to the original BBP model\,\cite{BBP}.
These complex phases will lead to a complex CKM matrix $\Vckm$ which diagonalizes
\eq{
\Vckm^\dag M_dM_d^\dag\Vckm=\diag(m_d^2,m_s^2,m_b^2)\,.
}
At the same time, spontaneous CP4 breaking also leads to a mass-squared splitting of the heavy quarks in \eqref{MD} proportional to $v_S^2$.
In the regime $\mu_D\lesssim M_{DS}$ we are interested in, this mass splitting is at least of the order of $\mu_D$.
This regime also means that the mixing matrix $U_R$ is not hierarchical and generically contains order one mixing angles and phases.

The presence of vector-like quarks $\cD_a$ implies the existence of FCNC interactions through $Z$
mediation which are however suppressed by the ratio between the masses of SM quarks and heavy quarks\,\cite{aguila}. 
Other effects such as electroweak precision observables or deviation of SM couplings are also suppressed for heavy quarks\,\cite{Aguilar-Saavedra:2013qpa}.
Current experimental searches at colliders constrains these heavy quarks to be heavier than the TeV scale\,\cite{CMS,ATLAS}.

\section{Loop corrections to $\btheta$}
\label{sec:loop}

As the model implements spontaneous CP violation and satisfies the Barr criteria\,\cite{barr,barr.2}, both the contributions coming from QCD and from the electroweak sector to the CP violating parameter $\btheta$ vanish at tree-level.
Higher order finite corrections are calculable and we will quantify these corrections in the following.
We will conclude that the one-loop contribution vanishes and nonzero contributions arise only at two-loops.

If we denote by $m_R$ the generic tree-level quark mass matrix in the basis $\bar{f}_L f_R$, it receives corrections at higher order as $m_R-\delta m_R$. This correction, if complex, will lead to a correction to $\btheta$ of the form
\eqali{
\label{delta.theta}
\delta\btheta&=\arg[\det(m_R-\delta m_R)]-\arg[\det(m_R)]
\cr
&\approx -\im[\Tr(m_R^{-1}\delta m_R)]
\,.
}
Only the corrections coming from the Yukawa interactions \eqref{yukawa} lead to a potentially complex contribution at one-loop\,\cite{BBP}.
Using dimensional regularization and \msbar\ we find at one-loop
\eq{
\label{delta.theta:1-l}
\delta\btheta=+\sum_{\varphi=h,s,A}\frac{1}{16\pi^2}
\im\{\Tr[m_R^{-1}Y^{\varphi}_R m_R^\dag I_L^{\varphi}Y^{\varphi}_R ]\}\,,
}
where 
\eq{
\label{ln}
I_L^{\varphi}=\int_0^1dx\log\left[\frac{x^2m_R m_R^\dag+(1-x)m^2_\varphi}{\mu^2}\right]
\,
}
is a loop function that depends on the possibly non-diagonal fermion mass matrix $m_R m_R^\dag$.
The calculation of the correction \eqref{delta.theta:1-l}, which is detailed in appendix \ref{ap:self}, is the same as found in Ref.\,\cite{BBP} if we ignore the renormalization scale $\mu$.
The Yukawa couplings are defined by
\eq{
-\lag\supset \sum_{\varphi}(\bar{f}_LY_R^\varphi f_R+\bar{f}_RY_R^{\varphi\dag} f_L)\varphi\,.
}
In appendix \ref{ap:finite} we explicitly show that the $1/\eps$ part of \eqref{delta.theta:1-l} vanishes at one-loop.
Hence these corrections are finite as expected.
This property also implies that $\delta\btheta$ is invariant by shifts of the subtraction point $\mu\to \mu+\delta \mu$ in \eqref{ln} so that we can use masses and Yukawa couplings at any suitable renormalization scale.

Ignoring the mass matrix from the up-type quarks, which is real, we can focus on the contribution for $m_R=\cM_D$.
Then the correction can be written as
\eq{
\label{delta.theta:1-l:2}
\delta\btheta=\mathop{\sum_{\varphi=h,s,A}}_{f}\frac{1}{16\pi^2}
\im[U_L^\dag\hat{C}^{\varphi\varphi}U_L]_{ff}
\int_0^1 dx\log\left[\frac{x^2m^2_f+(1-x)m^2_\varphi}{\mu^2}\right]
\,,
}
where the index $f$ runs through the charge $-1/3$ quark mass eigenstates and
\eq{
\hat{C}^{\varphi\varphi}\equiv Y^{\varphi}_R \cM_D^{-1}Y^{\varphi}_R \cM_D^\dag
}
is a matrix that transforms as $\bar{f}_{iL}f_{jL}$ in flavour space.
The coefficients $\hat{C}$ are defined for Yukawa couplings $Y^{\varphi}_R$ for mass eigenstates $\varphi\in \{h,s,A\}$.
The symmetry structure is more evident in the initial symmetry basis 
$\rho\in \{S_0,S_1,S_2\}$ connected through the mixing \eqref{mixing} which we write here as
\eq{
\rho=\sum_{\varphi} R_{\rho\varphi}\varphi\,,
}
although $R$ contains no mixing between $A$ and the CP even scalars.

The matrix $\hat{C}^{\varphi\varphi}$ can be written in terms of $C^{\rho\rho'}$ in the symmetry basis as
\eq{
\hat{C}^{\varphi\varphi}=\sum_{\rho\rho'}C^{\rho\rho'}R_{\rho\varphi}R_{\rho'\varphi}\,,
}
where 
\eq{
C^{\rho\rho'}
=Y^{\rho}_R \cM_D^{-1}Y^{\rho'}_R \cM_D^\dag\,.
}
The Yukawa couplings in the symmetry basis can be easily extracted from
\eq{
\sum_{\rho}Y^\rho_R\,\rho=
\mtrx{\frac{1}{v}\mu_d S_0 & 0\cr 
\frac{1}{v_S}(M_{DS}S_1+M_FS_2) & 0
}\,,
}
where we have defined $M_F=\frac{v_S}{\sqrt{2}}\,i(F-\bar{F})$.
So we can write generically 
\eq{
\label{Y:rho}
Y^\rho_R=\mtrx{A^\rho & 0\cr B^\rho& 0}\,,
}
where we note that among $A^\rho$ only $A^{S_0}$ is nonzero.

Explicit calculation leads to 
\eq{
C^{\rho\rho'}
=\mtrx{
A^\rho \mu_d^{-1}A^{\rho'}\mu_d^\dag & A^\rho \mu_d^{-1}A^{\rho'}M_{DS}^\dag
\cr
B^\rho\mu_d^{-1}A^{\rho'}\mu_d^\dag & B^\rho\mu_d^{-1}A^{\rho'}M_{DS}^\dag
}\,.
}
Since all entries involves $A^{\rho'}$, it is clear that a nonvanishing matrix requires $\rho'=S_0$ for which $\mu_d^{-1}A^{\rho'}=v^{-1}\id_3$.
Therefore,
\subeqali[C:rho.rho]{
C^{S_0S_0}&=
\frac{1}{v^2}\mtrx{\mu_d\mu_d^\dag & \mu_d M_{DS}^\dag
\cr 0 & 0
}
=\frac{1}{v^2}\mtrx{\id_3&\cr &0_2}\cM_D\cM_D^\dag\,,
\\
C^{S_1S_0}&=
\frac{1}{vv_S}\mtrx{
0 & 0\cr
M_{DS}\mu_d^\dag & H_D-\mu_D^2\id_2
}
=\frac{1}{vv_S}\mtrx{0_3&\cr &\id_2}\big(\cM_D\cM_D^\dag-\mu_D^2\id_5\big)\,,
\\
\label{C:S2S0}
C^{S_2S_0}&=
\frac{1}{vv_S}\mtrx{
0 & 0\cr
M_{F}\mu_d^\dag & M_{F} M_{DS}^\dag
}\,.
}
One can check that the diagonal elements of $U_L^\dag C^{\rho\rho'}U_L$ are real for $(\rho,\rho')=(S_0,S_0)$ or $(S_1,S_0)$.
For example,
\eq{
U_L^\dag C^{S_0S_0}U_L
=\frac{1}{v^2}U_L^\dag \mtrx{\id_3&\cr &0_2} U_L
\mtrx{\hat{M}_d^2 & 0\cr 0 & \hat{M}_D^2
}\,,
}
where the hatted matrices denote the diagonalized masses in \eqref{MD}.
The element $C^{S_2S_0}$ leads to a potentially complex contribution but the absence 
of mixing between $S_2$-$S_0$, i.e., $R_{S_2\varphi}R_{S_0\varphi}=0$ for all $\varphi$, makes all contributions to \eqref{delta.theta:1-l:2} vanish and there is no correction to $\btheta$ at one-loop.
This calculation is exact with respect to the mixing matrix $U_L$.
Therefore, our model predicts a correction to $\btheta$ only at two-loops, a feature that improves over the original BBP model.

\begin{figure}[h]
\includegraphics[scale=1]{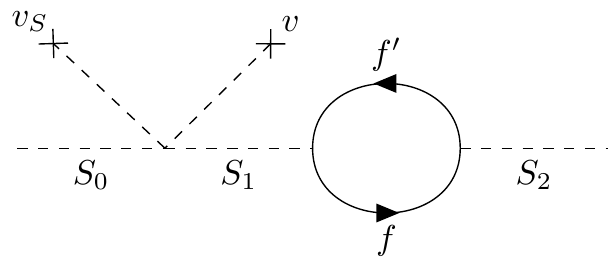}
\caption{\label{fig.1}
One-loop mixing of $h$ and $A$.
}
\end{figure}

In order to estimate the two-loop contribution to $\btheta$, we notice that the mixing between $S_{0}{-}S_{2}$ can be induced at one-loop level as shown in Fig.\,\ref{fig.1}, leading to a mixing angle of the order of
\eq{
\delta\alpha=|R_{S_2 h}|\sim |R_{S_0A}|\approx 
\frac{v}{v_S}\frac{\lambda_{\phi S}F^2}{16\pi^2}\,,
}
where $\lambda_{\phi S}$ is the Higgs portal coupling of $S$ and $F$ here denote a generic combination of $F_{aj}$.
The dominant contribution to the $h$-$A$ self-energy comes from the chirality flipping part of the heavy fermion propagators with insertion of $m_{\cD_a}$.
This mixing will induce a contribution to \eqref{delta.theta:1-l:2} coming dominantly from the lower-right block of \eqref{C:S2S0} because the diagonalization matrix \eqref{UL} is approximately block diagonal.
We arrive at
\eq{
\label{deta.theta:2-l}
\delta\btheta
\approx -\frac{1}{16\pi^2}\frac{v_S}{2v}
\mathop{\sum_{\varphi=h,A}}_{a=1,2}
\re[V_D^\dag(F-\bar{F})(F+\bar{F})^\dag V_D]_{aa}
R_{S_0\varphi}R_{S_2\varphi}I(M^2_{\cD_a},m^2_{\varphi})\,,
}
which can be estimated as
\eq{
\delta\btheta\approx 
\frac{F^2}{16\pi^2}\frac{v_S}{v}\times \delta\alpha=
\frac{\lambda_{\phi S}F^4}{(16\pi^2)^2}
\sim 4\times 10^{-5}\lambda_{\phi S}F^4\,.
}
The absence of the neutron eletric dipole moment constrains $\btheta \lesssim 3.0\times10^{-10}$\,\cite{EDMreview,nEDM:exp} so we just need a moderately small Yukawa coupling of the order $F\sim 0.05\lambda_{\phi S}^{-1/4}$ 
independently of the scale $v_S$ which contrasts with the model in \cite{BBP}.
The function $I(m_1^2,m_2^2)$ is the integral in \eqref{delta.theta:1-l:2},
a dimensionless slowly varying function which is order one for a wide range of values and can be written as $B(m_1^2,m_2^2,m_1^2)$ in terms of the $B$ function of Passarino and Veltman\,\cite{passarino}.
Explicit forms and asymptotic values are shown in appendix \ref{ap:I}.

Finally, we can see that the expression in \eqref{deta.theta:2-l} vanishes if $\cD_a$ are degenerate because $\Tr[(F-\bar{F})(F+\bar{F})^\dag]=0$.
In fact, the mass splitting of $\cD_a$ only arises as a consequence of the spontaneous breaking of CP4 through $M_{DS}$.
So, for a small mass splitting, $\delta\btheta$ is proportional to this mass splitting.

Considering other contributions, few comments are in order:
\begin{itemize}
\item The key property in our model that guarantees $\btheta=0$ at one-loop is the automatic conservation at tree-level of usual CP, $S_2\to -S_2$, in the scalar potential \eqref{potential} as a consequence of CP4.
This means that the mixing between CP even and CP odd scalars, $S_1$-$S_2$ mixing, arises only at one-loop through the graph in Fig.\,\ref{fig.1}.

\item The symmetry $S_2\to -S_2$ is not spontaneously broken at tree-level as $\aver{S_2}=0$, and it is only broken by $F-\bar{F}$ in the Yukawa couplings.

\item From rough estimates of representative graphs, a net complex contribution requires the interference between the Yukawa coupling of $S_1$ with that of $S_2$ and nonzero vevs for them.
Thus $\aver{S_2}=0$ leads to the vanishing of the estimates of Ref.\,\cite{dine} to $\btheta$ from (a) the one-loop threshold effect to the low-energy Yukawa coupling of $d$-type quarks (essentially the BBP contribution that we calculated) and (b) the two-loop complex contribution to the heavy quark mass matrix ($\mu_D$) from the dead-duck type diagram. 

\item Two-loop contributions to $\btheta$ unsuppressed by $F,\bar{F}$ (but suppressed by SM yukawas) will likely arise by using an effective theory at an intermediate energy scale between the electroweak scale and that of the heavy quarks when the other degrees of freedom are much heavier
(by, e.g., suppressing $F\ll 1$ so that $v_S\gg M_{\cD_a}$).
In the mass basis for these heavy quarks, all CP violation comes from the Yukawa interactions between light SM $d$-type quarks and the Higgs.
Then corrections to $\btheta$ arise only from two-loop corrections in these complex Yukawa couplings\,\cite{nelson:calc}.

\end{itemize}

\section{Conclusions}
\label{sec:conclu}

A very simple model is presented where CP violation in Nature is spontaneous in origin and then all CP violation effects are calculable. 
While easily accommodating the observed CP phase residing in the CKM mechanism, the $\btheta$ parameter of the QCD vacuum structure vanishes not only at tree-level --- the Nelson-Barr mechanism --- but also at one-loop level due to the imposition of a nonconventional CP symmetry of order 4, also known as CP4, on the fermion and scalar sector beyond the SM.
Thus the strong CP problem is solved with only moderate Yukawa couplings that couple the mediator heavy quarks, SM quarks and heavy scalars.
The field content of the SM is enlarged by adding just two vector-like $d$-type quarks and one complex singlet scalar.
No other symmetry is necessary besides the nonconventional CP.
Therefore, this model improves on the minimal model of Bento-Branco-Parada\,\cite{BBP} on two aspects: (a) there is no unrelated $\ZZ_2$ symmetry and (b) corrections to $\btheta$ arise only at two loops.%
\footnote{%
A similarly simple model with corrections only at two-loops is shown in \cite{lavoura:soft}.
However, an ad hoc $\ZZ_2$ is still necessary and making the soft CP breaking spontaneous would introduce more fields. The authors thank L.~Lavoura for pointing out his work.
}

\acknowledgements

The authors thank João Silva, Igor Ivanov and Andreas Trautner for helpful discussions.
C.C.N.\ acknowledges partial support by brazilian Fapesp, grant 2014/19164-6, and
CNPq, grant 308578/2016-3. A.L.C.\ acknowledges financial support by the Coordenação de Aperfeiçoamento de Pessoal de Nível Superior - Brasil (CAPES) - Finance Code 001.

\appendix
\section{Self-energy}
\label{ap:self}

Here we provide further details on the calculation of the fermion self-energy at one-loop level. Following similar arguments to those presented at \cite{BBP}, only the contribution due to the exchange of a scalar field will be relevant to $\bar{\theta}$. Therefore, we will only need Yukawa couplings which can be generally defined by
\eq{
-\lag=\sum_{\varphi}\bar{f}_iY^{\varphi}_{ij}f_j\varphi\,,
}
where $\varphi$ are real scalar fields and $Y^{\varphi}_{ij}$ may contain $\gamma_5$ matrices in Dirac space. It is hermitean in the sense that $\gamma_0(Y^{\varphi})^\dag\gamma_0=Y^{\varphi}$. 

We are mainly interested in the one-loop contribution to $\bar{f}_if_j$. Assuming $\varphi$ are mass eigenstates, the amputated diagram, which also contains an internal fermion $f_k$, can be written as 
\eq{
i\Sigma_{ij}(\slashed{p})=
\int \frac{d^{4}k}{(2\pi)^4}
\frac{Y^{\varphi}_{ik}(\slashed{k}+m_k)Y^{\varphi}_{kj}}{(k^2-m^2_k)[(k-p)^2-M_{\varphi}^2]},
}
where a sum on the scalars $\varphi$ is implicitly assumed. Although the contribution to $\bar{\theta}$ is finite, a regularization is needed in order to deal with intermediate steps of the calculation. In this work, we will choose to adopt the dimensional regularization scheme, which yields
\eq{
i\Sigma_{ij}(\slashed{p})=
\frac{i}{16\pi^2}
Y^{\varphi}_{ik}\int_0^1dx\big[\slashed{p}(1-x)+m_k\big]
\left\{\frac{1}{\varepsilon}
+\ln\left[\frac{\tilde{\mu}^2}{xm^2_k+(1-x)M^2_{\varphi}-x(1-x)p^2}\right]
\right\}
Y^{\varphi}_{kj}
\,.
}

The calculation can be performed in a generic fermion mass basis as well,
in which we can decompose
\eq{
\label{m:LR}
m=m_RR+m_L L\,,\quad Y^{\varphi}=Y^{\varphi}_{R}R+Y^{\varphi}_{L}L\,,
}
where $R=(1+\gamma_5)/2$ and $L=(1-\gamma_5)/2$. Notice that the hermitean condition 
implies $Y^{\varphi}_{L}=Y^{{\varphi}\dag}_R$, and $m_L=m_R^\dag$. Therefore, the self-energy can be expressed as
\eq{
i\Sigma_{ij}(\slashed{p})
=i\left[\slashed{p}R\Sigma^{R}_{ij}(p^2) + \slashed{p}L\Sigma_{ij}^{L}(p^2) + \Sigma_{ij}^{m}(p^2)\right],}
where we defined
\eqali{
i\Sigma^{R}_{ij}(p^2) &= \frac{i}{16\pi^2}(Y^{\varphi}_{L})_{ik}\int_0^1dx(1-x)X^{\varphi}_{kl}(Y^{\varphi}_{R})_{lj},\quad 
\Sigma^{L}(p^2) = \Sigma^{R}(p^2)\big|_{(L,R)\rightarrow (R,L)},\\
i\Sigma_{ij}^{m}(p^2) &= \frac{i}{16\pi^2}(Y^{\varphi}_{R})_{ik} (m_{R}^{\dag})_{kl}\int_0^1dx X^\varphi_{ll'} (Y^{\varphi}_{R})_{l'j} + (R\rightarrow L),\\
X^\varphi_{kl} &=
\frac{1}{\varepsilon}\delta_{kl}
+\left\{\ln\left[\frac{\tilde{\mu}^2}{x\,m_{R}m_{R}^{\dagger}+(1-x)M^2_{\varphi}-x(1-x)p^2}\right]\right\}_{kl}.
}
As it stands, the self-energy still contains a divergent piece. We show in appendix \ref{ap:finite} that this term will eventually drop in the calculation of $\delta\bar{\theta}$, however, for definiteness, one can adopt a subtraction scheme such as the \msbar\ and explictly remove the $\varepsilon^{-1}$ term.

Finally, regarding $\delta\bar{\theta}$, the relevant quantity is the radiative correction to the mass matrix, namely $m_{R} - \delta m_{R}$. To obtain such, one needs to find the position of the poles of the corrected fermion propagator
\eq{
\Delta = \left[\slashed{p} - m +\Sigma(p^2)\right]^{-1}.
}
Using the chiral decomposition and defining
\eq{
\label{eq:eff}
\Sigma^{\rm eff}(p^2) = \ums{2}m\left(L\Sigma^{L}(p^2) + R\Sigma^{R}(p^2)\right) + \ums{2}\left(R\Sigma^{L}(p^2) + L\Sigma^{R}(p^2)\right)m + \Sigma^{m}(p^2),
}
one obtains, to first order on the corrections,
\eqali{
\Delta &= \left[\frac{(\slashed{p}-m)}{2}\left(L\Sigma^{L}(p^2) + R\Sigma^{R}(p^2)\right)+\left(R\Sigma^{L}(p^2) + L\Sigma^{R}(p^2)\right)\frac{(\slashed{p}-m)}{2}+\Sigma^{\rm eff}(p^2)\right]^{-1},\\
&\approx \bigg[1-\ums{2}\left(R\Sigma^{L}(p^2) + L\Sigma^{R}(p^2)\right)\bigg]^{-1}\left[\slashed{p} - m +\Sigma^{\rm eff}(p^2)\right]^{-1}\bigg[1-\ums{2}\left(L\Sigma^{L}(p^2) + R\Sigma^{R}(p^2)\right)\bigg]^{-1}.
}
Therefore, the position of the poles can be found with knowledge only of $\Sigma^{\rm eff}(p^2)$, and the end result is
\eq{
\label{eq:dmr}
\delta m_{R} = \sum_{\varphi=h,s,A}\bigg(\ums{2}m_{R}\Sigma^{R} + \ums{2}\Sigma^{L}m_{R} + \Sigma_{R}^{m}\bigg),
} 
where we consider only the R projection of Eq.\,\eqref{eq:eff}, the sum on the scalars is explicitly introduced and $p^2 = m_{R} m_{R}^{\dagger}$ at leading order. 
Finally, since both $\Sigma^{L}$ and $\Sigma^{R}$ are hermitean, only the last term of Eq.\,\eqref{eq:dmr} contributes to $\delta\btheta$.

\section{One-loop correction to $\btheta$ is finite}
\label{ap:finite}

Here we show that the one-loop correction in \eqref{delta.theta:1-l} is finite by explicitly retaining the $1/\eps$ terms of dimensional regularization.
The correction is given by
\eq{
\delta\btheta=-\im\Tr[m_R^{-1}\delta m_R]\,,
}
where
\eq{
\delta m_R=\Sigma^m_{R}+\ums{2}\Sigma^L m_R+\ums{2}m_R\Sigma^R\,.
}

The coefficient of $(16\pi^2\eps)^{-1}$ in $\delta m_R$ coming from generic Yukawa couplings is 
\eq{
\label{1/eps}
\delta m_R\big|_{1/\eps}=
\sum_{\rho=\text{scalars}}\left(
Y^{\rho}_R  m_R^\dag Y^{\rho}_R +\ums{4}Y^{\rho}_R Y^{\rho\dag}_R m_R+\ums{4}m_RY^{\rho\dag}_R Y^{\rho}_R 
\right)\,,
}
where $Y^\varphi_R$ is written in a generic basis for fermions and for scalars.
We can see that only the first term of \eqref{1/eps} leads to a potentially complex contribution as
\eq{
\delta\btheta\big|_{1/\eps}=-\sum_\rho\im\Tr[m_R^{-1}Y^{\rho}_R  m_R^\dag Y^{\rho}_R ]\,.
}

For Branco-Bento-Parada type models, we can use \eqref{Y:rho}
with real and nonzero $A^\rho $ for $\rho =\sqrt{2}\re\phi^0$ of the Higgs doublet and 
zero otherwise.
In contrast, $B^\rho $ is nonzero and generically complex only for the scalars beyond the SM.
If we use the mass matrix structure in \eqref{mass.matrix:d,D} for $m_R$ we arrive at
\eq{
\label{delta.theta:1/eps:2}
\delta\theta\big|_{1/\eps}=-\sum_\rho\im\Tr[\mu_d^{-1}A^\rho (\mu_d^\dag A^\rho +M_{DS}^\dag B^\rho )]\,.
}
Only $M_{DS}$ and $B^\rho $ can be complex but $A^\rho $ and $B^\rho $ cannot be simultaneously nonzero for the same scalar $\rho$ so \eqref{delta.theta:1/eps:2} vanishes and hence the one-loop contribution to $\delta\btheta$ is finite.

\section{Loop function}
\label{ap:I}

The loop function defined by the integral in \eqref{delta.theta:1-l:2} can be written in different forms as
\eqali{
I(m^2_f,m^2_\varphi)&\equiv
\int_0^1 dx\log\left[\frac{x^2m^2_f+(1-x)m^2_\varphi}{\mu^2}\right]\,,
\cr
&=B(m_f^2,m_\varphi^2,m_f^2)\,,
\cr
&=\log\bigg[\frac{m^2_f}{\mu^2}\bigg]+h(m^2_\varphi/m^2_f)\,,
}
where $B$ is the Passarino-Veltman function\,\cite{passarino}:
\eq{
\label{passarino}
B(m_1^2,m_2^2,s) \equiv  \int^1_0 \,dx\, \log \left[ \frac{x\,m_{1}^2 + (1 - x)\,m_{2}^2 - x(1 - x) s}{\mu^2}\right].
}
In the last form, the function $h$ is 
\eqali{
h(z)&\equiv\int_0^1dx\log\left[x^2+(1-x)z\right]
\cr
&=\ums{2}z\log(z)-2+z\sqrt{\frac{4}{z}-1}\arcsin\sqrt{1-\frac{z}{4}}\,,\quad \text{ if $z\le 4$}\,,
\cr
&=\ums{2}z\log(z)-2-z\sqrt{1-\frac{4}{z}}\operatorname{arcsinh}\sqrt{\frac{z}{4}-1}\,,\quad \text{ if $z>4$}\,.
}
The function in Ref.\,\cite{BBP} coincides with $h(z)+2$ which is always non-negative.
The function $h(z)$ is a monotonically increasing function with the asymptotic behaviour
\eqali{
h(z)&\approx -2+\pi\sqrt{z}\,,\quad \text{for $z\ll 1$,}
\cr
h(z)&\approx\log(z)-1\,,\quad \text{for $z\gg 1$.}
}
Therefore, this function varies very slowly and is typically of order one for a wide range of values for $z$.
We can also write $h(z)=B(1,z,1)$.


\end{document}